\documentclass[sigconf]{acmart}
\usepackage{graphicx}

\usepackage{caption}
\usepackage{float}
\AtBeginDocument{%
  }

\setcopyright{acmlicensed}
\copyrightyear{2025}
\acmYear{2025}
\acmConference[TSMO at KDD]{Workshop on Two-sided Marketplace Optimization: Search, Discovery, Matching, Pricing \& Growth}{August 4th, 2025}{Toronto, Canada}
\begin{document}

\title{Personalized Recommendation of Dish and Restaurant Collections on iFood}

\author{Fernando F. Granado}
\authornote{All authors contributed equally to this research.}
\email{fernando.granado@ifood.com.br}
\affiliation{%
  \institution{iFood}
  % \city{São Paulo}
  \state{São Paulo}
  \country{Brazil}
}

\author{Davi A. Bezerra}
% \authornote{All authors contributed equally to this research.}
\authornotemark[1]
\email{davi.bezerra@ifood.com.br}
\affiliation{%
  \institution{iFood}
  % \city{São Paulo}
  \state{São Paulo}
  \country{Brazil}
}

\author{Iuri Queiroz}
\authornotemark[1]
% \authornote{All authors contributed equally to this research.}
\email{iuri.queiroz@ifood.com.br}
\affiliation{%
  \institution{iFood}
  % \city{São Paulo}
  \state{São Paulo}
  \country{Brazil}
}

\author{Nathan Oliveira}
\authornotemark[1]
% \authornote{All authors contributed equally to this research.}
\email{nathan.oliveira@ifood.com.br}
\affiliation{%
  \institution{iFood}
  % \city{São Paulo}
  \state{São Paulo}
  \country{Brazil}
}

\author{Pedro Fernandes}
\authornotemark[1]
% \authornote{All authors contributed equally to this research.}
\email{fernandes.pedro@ifood.com.br}
\affiliation{%
  \institution{iFood}
  % \city{São Paulo}
  \state{São Paulo}
  \country{Brazil}
}

\author{Bruno Schock}
\authornotemark[1]
% \authornote{All authors contributed equally to this research.}
\email{bruno.schock@ifood.com.br}
\affiliation{%
  \institution{iFood}
  % \city{São Paulo}
  \state{São Paulo}
  \country{Brazil}
}


\begin{abstract}
Food delivery platforms face the challenge of helping users navigate vast catalogs of restaurants and dishes to find meals they truly enjoy. This paper presents RED, an automated recommendation system designed for iFood, Latin America's largest on-demand food delivery platform, to personalize the selection of curated food collections displayed to millions of users.
Our approach employs a LightGBM classifier that scores collections based on three feature groups: collection characteristics, user-collection similarity, and contextual information. To address the cold-start problem of recommending newly created collections, we develop content-based representations using item embeddings and implement monotonicity constraints to improve generalization. We tackle data scarcity by bootstrapping from category carousel interactions and address visibility bias through unbiased sampling of impressions and purchases in production.
The system demonstrates significant real-world impact through extensive A/B testing with 5-10\% of iFood's user base. Online results of our A/B tests add up to 97\% improvement in Card Conversion Rate and 1.4\% increase in overall App Conversion Rate compared to popularity-based baselines. Notably, our offline accuracy metrics strongly correlate with online performance, enabling reliable impact prediction before deployment. To our knowledge, this is the first work to detail large-scale recommendation of curated food collections in a dynamic commercial environment.
\end{abstract}





\maketitle

\section{INTRODUCTION}
% \subsection{Motivation}

iFood is Latin America's largest on-demand food delivery platform,  
connecting millions of users to hundreds of thousands of restaurants in Brazil. 
As user expectations for fast, convenient, and personalized experiences continue to rise, 
one of iFood's greatest challenges is helping diners cut through an overwhelming array 
of menu options to find meals they will truly enjoy. Addressing this challenge is essential for driving customer satisfaction, increasing engagement, and fostering long-term loyalty.

Recommendation systems play a central role in meeting this challenge 
by delivering tailored meal suggestions that guide users through iFood’s vast catalog. 
Rather than merely proposing restaurants based on location or popularity, 
these systems dive deeper into individual dish preferences—powering features ranging from home-feed carousels filled with meals predicted to match a user’s flavor profile to personalized rankings of dishes and restaurants. 
By continuously learning from past orders, ratings, and contextual signals (e.g., time of day or ongoing promotions), 
the goal is to boost both user satisfaction and conversion rates by placing the most appealing options front and center.

However, crafting truly personalized food recommendations is far from straightforward, 
because food preference is deeply individual and shaped by a complex web of influences. What one person finds irresistible, another may find unappealing, 
as choices are driven not only by inherent taste sensitivities 
and sensory experiences but also by cultural upbringing, dietary restrictions, 
and broader socio-economic conditions \cite{leng2017}. Even the same person's preferences can change rapidly—tolerating one hour of delivery time when logging in at 11:00 a.m. but being extremely time-sensitive when logging in at 11:50 a.m. on working days, or having very different cuisine preferences depending on the time of day and social context. Moreover, factors such as the restaurant's 
constantly changing offerings, regional cuisine diversity, 
and even time-sensitive promotions all complicate the task of surfacing the right dishes at the right moment. Accounting for this rich diversity of drivers is therefore essential 
for any recommender system that aspires to deliver genuinely relevant meal suggestions.

Since 2019, iFood has been working on app personalization, initially deploying models based on matrix factorization and reinforcement learning techniques 
to recommend dishes, achieving significant improvements in click-through rates and session conversion. 
Despite these quantitative successes, qualitative customer feedback highlighted substantial shortcomings: 
users frequently ignored certain homepage components—particularly promotional banners—a phenomenon known as ``banner blindness.'' 
Users expressed feeling overwhelmed by advertisements, reporting minimal personalization. 
Although users acknowledged accurate item ordering reflecting their tastes, 
they emphasized the app's failure to address specific individual needs. 
This insight underscored the necessity of improving content contextualization and recommendation relevance for iFood users.

\subsection{Solution Architecture}

Through comprehensive analysis of iFood's content structure, 
we identified three strategic avenues for enhancing personalization:

\begin{enumerate}
\item \textbf{Semi-personalized Recommendations}: Segmenting customers based on consumption patterns and delivering manually curated content for each segment via the Content Management System (see Supplementary Information for details on the CMS).
\item \textbf{Curated Collection Recommendations}\footnote{A \textit{collection} is defined as a set of dishes or restaurants with a common theme, such as "Japanese Food" or "Free Delivery"}: Implementing models that evaluate and select the most relevant pre-created content snippets tailored to each user, including personalized text and layout adjustments.
\item \textbf{Personalized Collection Generation}: Utilizing AI-driven methods to dynamically assemble unique dish sequences for individual users, optimizing both content ordering and textual/visual presentation.
\end{enumerate}

\subsection{Previous Approaches and Current Direction}

To validate the first strategic opportunity—semi-personalized recommendations—we constructed a clustering model to segment users based on:

\begin{itemize}
\item The average FastText\footnote{https://fasttext.cc/docs/en/python-module.html} embedding of previously ordered items;
\item Average delivery times;
\item Average order values;
\item Total number of past orders.
\end{itemize}

Deploying tailored content for each user segment resulted in a measurable impact: a 1.06\% lift in homepage conversion and a 0.58\% increase in overall app conversion ($p < 0.05$), corresponding to hundreds of thousands of additional organic orders monthly. However, despite these positive outcomes, the manual content management introduced operational overhead, limiting scalability for broader deployment.

To overcome scalability constraints, \textbf{we introduced RED, a model that dynamically selects the most relevant curated collections for each user}, ensuring high scalability and adaptability to changing business requirements—as described in our second strategic opportunity. By leveraging this automated approach, we eliminated manual overhead and significantly enhanced personalization and user experience.

Beyond static carousels and banners, iFood's app design evolved to present dynamically personalized components on the homepage. We employed a learning-to-rank framework, treating horizontal lists (collections) on the homepage as an ``item or restaurant set'' whose combined utility required comprehensive scoring and optimization, similar to personalized rows of videos on streaming platforms. This methodological advancement enabled nuanced, real-time adaptation to individual user tastes and contexts.

\section{RELATED WORK}

Recommending collections of items has been studied in a variety of domains. 
Examples include supermarket basket prediction \cite{vanmaasakkers2023}, 
outfit recommendation in fashion—where outfits rather than individual 
garments are suggested \cite{ma2022}—reading-list generation \cite{gobinda2023}, 
and music playlist creation \cite{tomasi2023}. 
Each of these tasks highlights the importance of capturing 
relationships such as complementarity among items within a set.

In the context of recommending dishes or restaurants, additional challenges arise. 
Different regions exhibit varying dietary preferences and definitions of meals; 
consumption patterns change rapidly throughout the day; and, unlike typical e-commerce products, the availability of menu items is constrained to a very small region.

We aim for our recommendations to be accurate and reflect individual user preferences, 
while ensuring sufficient diversity to meet the full range of user interests, 
rather than focusing on a narrow subset. 
To this end, our approach builds on previous work that 
systematically investigates the relationship between recommendation 
accuracy and the overall diversity of recommended items \cite{adomavicius2012}.

\begin{figure}
    \centering
    \includegraphics[width=0.75\linewidth]{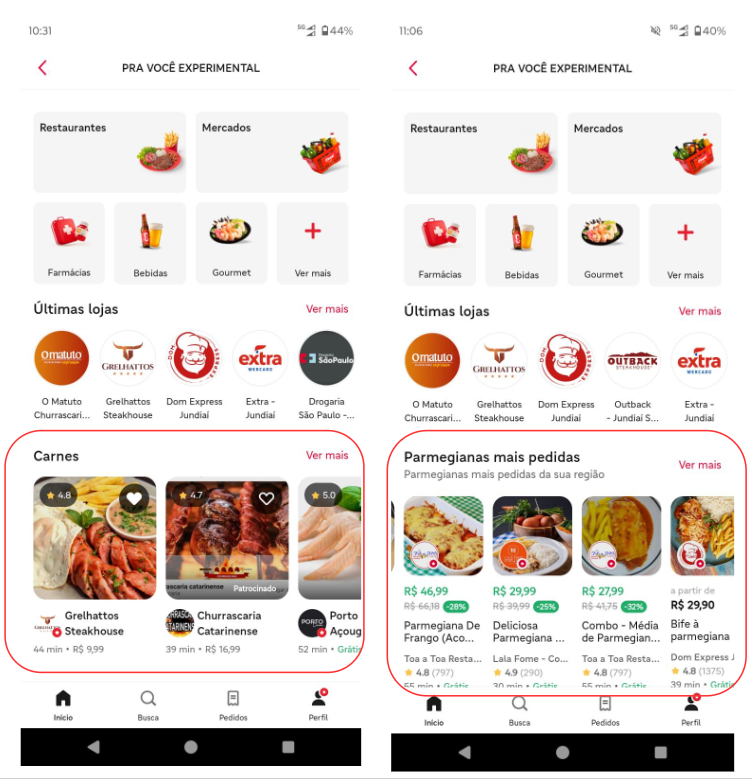}
    \caption{Personalized homepage recommendation layout showing horizontally scrollable rows of restaurant and dish collections.}
    \label{fig:home}
\end{figure}

Finally, to achieve our goal of facilitating app navigation, 
we draw inspiration from streaming platforms,
where recommendations appear on the homepage in a layout similar to ours. 
In our application, restaurants or dishes are presented in coherent horizontal 
rows that users can scroll through to discover more options, as in Figure \ref{fig:home}.
Consequently, a key element of our personalization strategy 
is the method we use to select which rows are displayed on the homepage.

\section{Problem Definition}\label{sec:problem_definition}

iFood has a team of experts responsible for creating curated collections of dishes and restaurants centered around specific themes. Given the large number of collections and the fact that most cater to very specific tastes, our system selects a small subset to display to each user on their homepage, based on a relevance score determined by our model.

As the team of experts creates new collections, the model has to generalize to collections not seen during training. 
Therefore, we cannot build a multi-class classifier corresponding to the different collections. 
Instead, we opt for creating content-based representations of collections, where each collection is described by its constituent items, restaurants, title, images, and other metadata and build 
a model that generates a score for each tuple ($user$, $collection$, $context$). 
More formally, the score $s$ is a function of the user $u$, the collection $c$, and the context $x$:

$$s = F(u, c, x)$$

With that framework, we can create a representation for the collections and generate scores as soon as they are created.

\section{MATERIALS AND METHODS}
This section describes the overall design of our collection recommendation pipeline, 
including the choice of scoring model, the feature engineering backbone, 
the dataset construction strategy, and the offline evaluation framework.

\subsection{Scoring Model}  
We employ a LightGBM‐based classifier \cite{ke2017lightgbm} trained on three sets of hand‐engineered features.  This choice trades off some end‐to‐end learning capacity in favor of greater interpretability, lighter inference cost, and rapid offline processing of all ($user$, $collection$, $context$) tuples.  Moreover, our feature definitions have been reused across multiple teams (e.g.\ in user profiles), enabling knowledge transfer to related projects. The model leverages the following feature groups:

\begin{itemize}
    \item \textbf{Collection}: captures characteristics of the collection definition and its contents. For instance, its popularity, whether it is a collection of dishes or restaurants, the percentage of orders with free delivery and how specific the collection is for the given meal shift (calculated as the percentage of orders made in that shift in relation to the total number of orders)
 
    \item \textbf{User-Collection Similarity}: captures how close the collection contents match characteristics of previous user's orders. For instance, the cosine similarity between user's and collection's embedding representation, the number of orders the user has placed in the collection's restaurants and a boolean variable indicating if the collection filters vegan content and the user is identified as vegan.
 
    \item \textbf{Context}: Features capturing characteristics of the moment the user is receiving the recommendation. For instance, the meal shift of the recommendation. 
\end{itemize}

\subsubsection{User embedding.}\label{sec:user_embedding} iFood has reusable item embeddings optimized for search retrieval, generated by a two-tower fashion model that approximates search terms to clicked items in vector space. In that context, we represent users as the embeddings of items that they previously bought. This encourages the model to recommend collections similar to the users' previous orders. However, instead of relying on a single item, we choose three items of distinct taxonomies to capture the users' diverse interests. The rule that selects these three items depends on the frequency with which the user buys from each taxonomy and item, as well as the order recency.

Furthermore, our analysis showed that the behavior of users in a meal shift provides little information about what she buys in different shifts. For instance, people who eat pizza at dinner do not necessarily eat pizza at breakfast. For that reason, we compute the users' embeddings independently for each meal shift.

Therefore, for each meal shift, we have up to three distinct items representing the user. We use each item independently to compute the similarity with the collections, leading to three embedding similarity features.

\subsubsection{Collection Embedding.}\label{sec:collection_embedding} To generate embeddings for the collections, we leverage a similar approach as for the users, reusing item embeddings. However, each collection is represented by a single embedding vector. For collection of dishes, we calculate the vector as the mean of the embeddings of their dishes and, for collections of restaurants, we first compute the embedding of the restaurants as the mean of their items, then aggregate again to compute the embedding of the collection.

\begin{figure}[h]
  \centering
  \includegraphics[width=\linewidth]{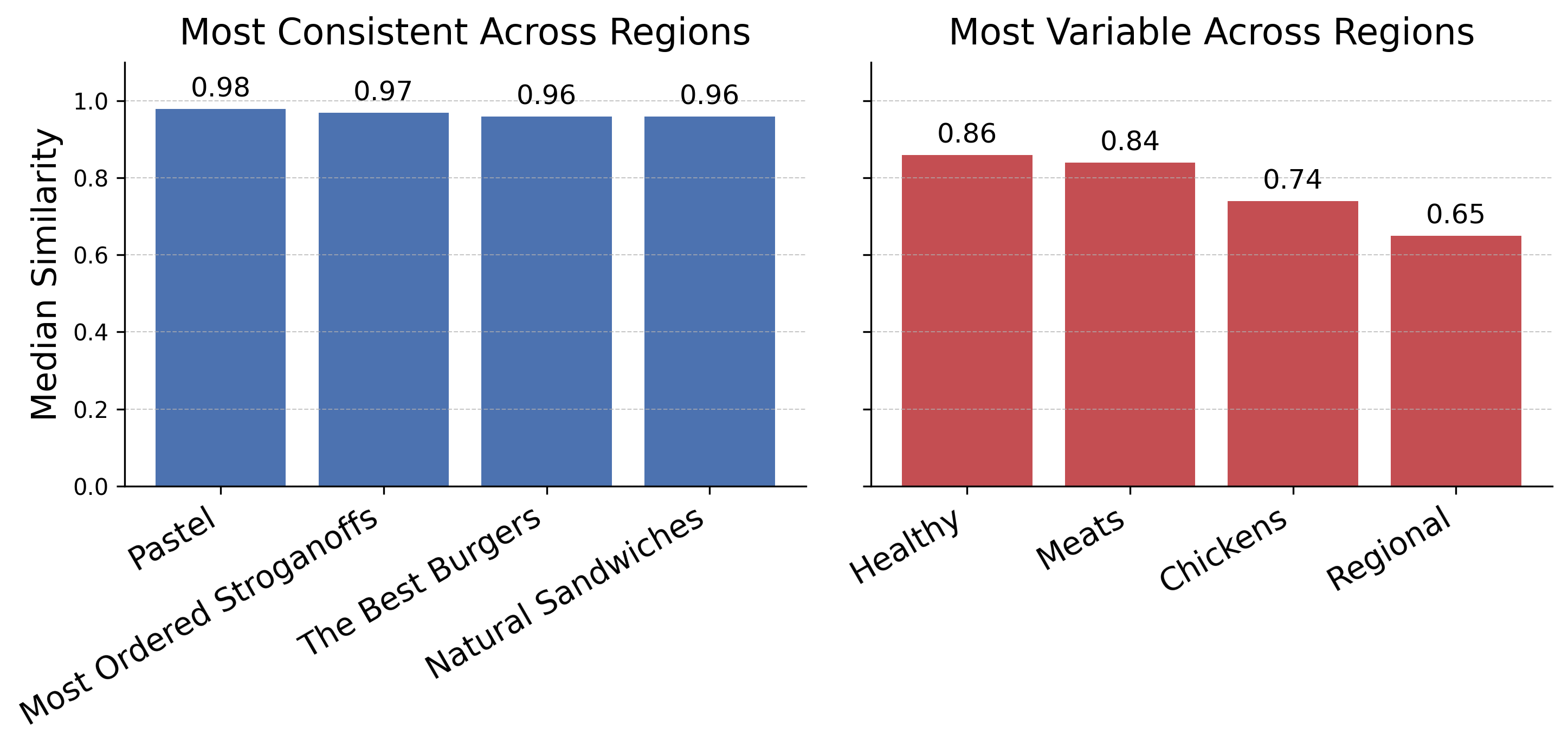}
  \caption{Median similarity between regional representations of collections and the unified representation across Brazil.}
  \label{fig:variability}
  \Description{.}
\end{figure}

Given that iFood operates across regions with diverse cultures, collection content can vary significantly by region. Figure~\ref{fig:variability} demonstrates this empirically, showing that some collections exhibit greater variability in their representations across regions. Our analysis found that using distinct embeddings for each state within the country yields a slight improvement in performance. However, we opted for a unified representation due to its simplicity.

\subsubsection{Controllability and Overfitting.}\label{sec:controlability} A key challenge is preventing overfitting to the relatively small set of training collections. Our model is trained on a set of \textasciitilde 100 collections and has to generalize to unseen ones. For the relatively small amount of samples, we observed that the model overfits to the collections in the training set, producing anomalous behaviors when applied to unseen data. For instance, it occasionally assigned a lower score when user–list similarity features increased or a higher score when the average delivery fee increased. To mitigate this, we impose monotonicity constraints on most collection and user-collection similarity features. This inductive bias both improves interpretability and enhances generalization to new lists.

\subsection{Dataset}
\label{sec:dataset}
The development of an effective collection recommendation system relies heavily on the availability of high-quality data that captures user preferences across different collection types and contexts. As our goal is to model user selection behavior among multiple collection options, it is crucial to identify data sources that provide clear and reliable signals of user intent. In this section, we describe the data sources, evolution, and methodology adopted for training the RED recommendation model, highlighting how we addressed the challenges of data scarcity and bias at different stages of system maturity.

\subsubsection{First Data Source Version.}\label{first_data} The RED recommendation model, as the first collection recommendation system on iFood, encountered a significant challenge during its initial development: the absence of live production data for the collections it was designed to recommend. This raised a critical question: How could we source data to train the early version of the model?
In building a collection recommendation model, it is essential to have a signal representing a user's preference among multiple collections. To address the lack of direct data, we explored existing app components that could provide a comparable signal for list recommendation. One such component was identified based on the following criteria:

\begin{itemize}
 \item High visibility within the app
 \item Presentation of diverse content to the same user 
 \item Possibility for implicit user feedback, such as clicks or purchases, to indicate collection preferences 
\end{itemize}

\begin{figure}[h]
  \centering
  \includegraphics[width=\linewidth]{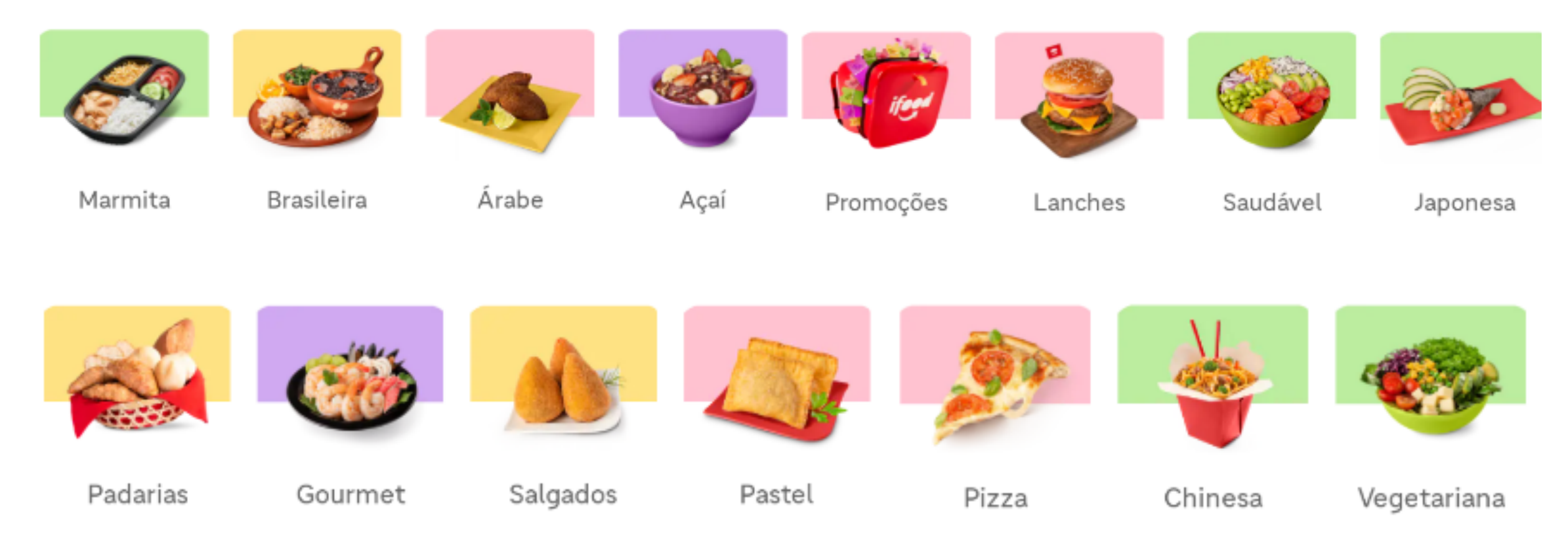}
  \caption{Rendered contents of the category carousel for a random user, used to train the initial version of our collection recommendation model.}
  \label{fig:carousel}
  \Description{.}
\end{figure}

We selected the category carousel, as depicted in Figure ~\ref{fig:carousel}, as a viable data source. This carousel displays icons representing various categories—such as ``Snacks,'' ``Brazilian,'' and ``Japanese''—and drives a substantial volume of monthly purchases. Using data derived from these categories enabled us to train the initial versions of the RED model. 

A comparison of signals for training revealed that purchase feedback was more effective than click feedback. While clicks reflected browsing behavior, purchases better represented user intent. Consequently, purchase feedback became our preferred signal for model training.

\subsubsection{Mature Data Source Version.}\label{sec:mature_data} The early dataset version come with inherent limitations:

\begin{enumerate}
 \item Restricted Diversity: The carousel collections are mostly limited by cousines, making it challenging for the model to optimize the recommendation of other themes, for example ``Shop with meal vouchers'', ``Best-sellers near you,'' or ``Affordable Lunch.'' 
 \item Visibility Bias: The ordering within the carousel affected data collection, often favoring certain categories. 
 \item Context Sensitivity: Users in different app sections exhibited varying interests, further complicating the training process. 
\end{enumerate}

\begin{table*}
  % \centering
  % \scriptsize
  % \setlength\tabcolsep{3pt}
  \caption{Online A/B test results of different RED versions. Values in bold achieved statistical significance (p < 0.05).}
  \label{tab:red_results}
  % \resizebox{\linewidth}{!}{%
    % \begin{tabular}{@{} ll p{3cm} ccc @{}}
    \begin{tabular}{cccccc}
      \toprule
      Control & Variant & Modification & CCR & HCR & ACR  \\ 
      \midrule
      Baseline & RED v1 
        & Model trained with carousel data
        & \textbf{+23.50\%}  & +0.57\%  & +0.02\% \\ 
      RED v1 & RED v2 
        & Monotonicity constraints + migrate to search embeddings
        & \textbf{+12.34\%}  & \textbf{+0.58\%}  & \textbf{+0.35\%} \\ 
      RED v2 & RED v3
        & Addition of collection popularity
        & \textbf{+11.51\%}  & \textbf{+0.97\%}  & +0.25\% \\ 
      RED v3 & RED v4
        & Training with sampled data 
        & \textbf{+18.32\%}  & \textbf{+0.72\%}  & +0.26\% \\ 
      RED v4 & RED v5
        & New similarity features 
        & \textbf{+6.30\%}   & \textbf{+0.88\%}  & +0.21\% \\ 
      \bottomrule
    \end{tabular}%
  % }
\end{table*}

To overcome the limitations of the early version, we transitioned to using impression and purchase data gathered directly from RED cards in production. We introduced a random sampling process for 0.5\% of displayed recommendations, collecting unbiased data essential for training subsequent, mature versions of the model. Although this approach reduced the raw data volume, the higher quality, unbiased data significantly improved training outcomes.

\subsubsection{Addressing Visibility Bias.} Another challenge concerning the training dataset relates to visibility bias. RED operates across various home sections, each with distinct traffic and conversion rates. Due to business constraints, the available collections vary across sections and user segmentation and change regularly. This leads to very different performances of collections depending on where they are displayed, distorting the model's perception of collection relevance. Consequently, the model tends to assign higher scores to collections that are displayed more frequently in high-conversion homes, even if they are not necessarily more relevant. 

To address this issue, the training dataset was generated following these rules: 
 
\begin{itemize}
 \item Only sessions that included a purchase on RED are considered for training. The purchased collection serves as the positive sample.
 \item The negative sample consists of another displayed list within the same homepage where the purchase occurred but without any purchase. 
\end{itemize}

\begin{figure}[h]
  \centering
  \includegraphics[width=\linewidth]{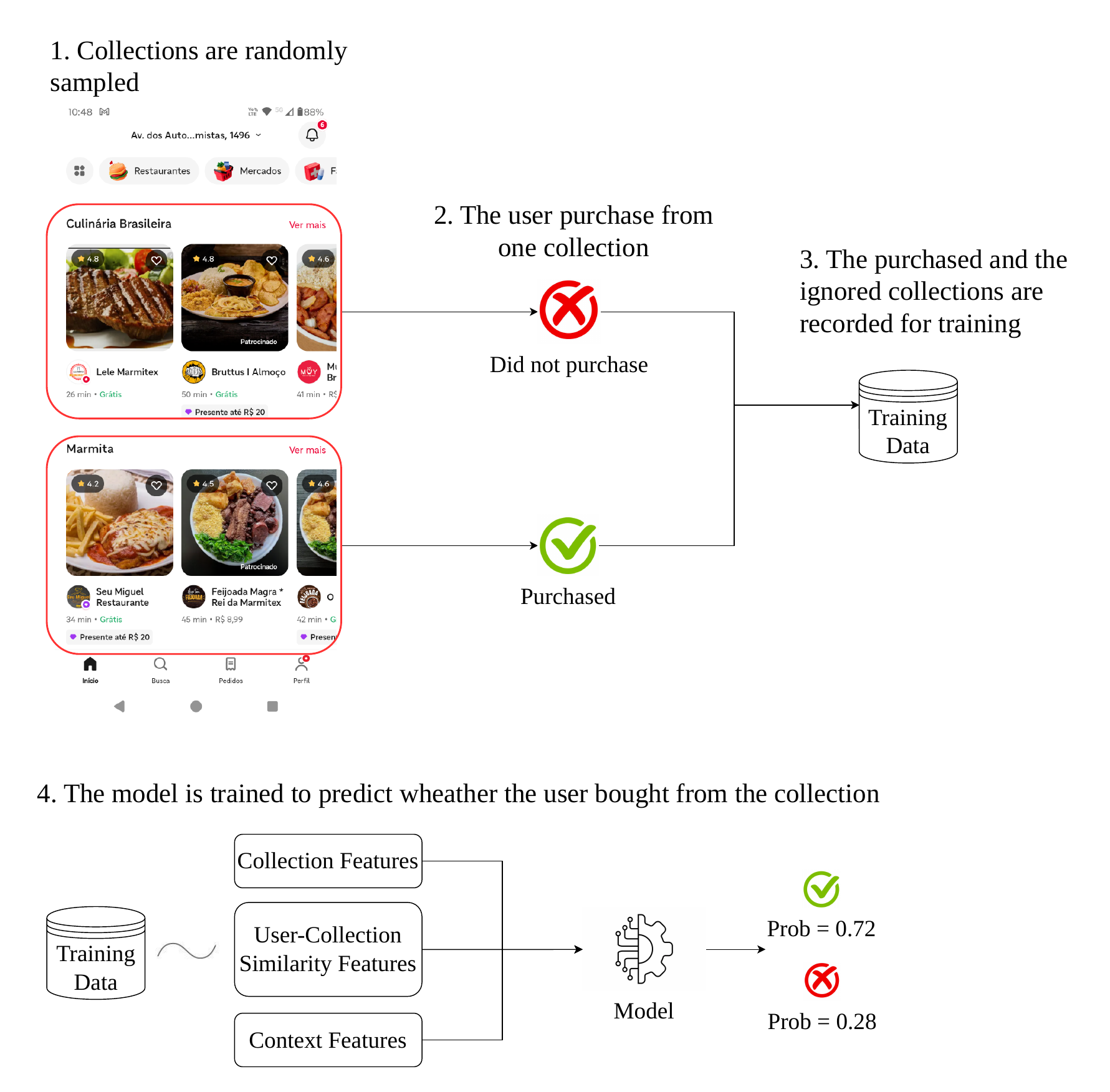}
  \caption{Overview of the methodology used to generate the training dataset. In 0.5\% of the recommendations, we show to the user a randomly sampled pair of collections drawn from the collections available for that space and user segment. We train the model in a balanced dataset, where the collection the user has bought from has positive label and the other one, with similar visibility properties, has negative label.}
  \label{fig:training_process}
  
  \Description{.}

\end{figure}

This ensures that the positive and negative samples in a given session always belong to the same home, offering similar visibility. Moreover, conversion rates across different homes are balanced, all containing 50\% positive and 50\% negative samples. The model was then trained to determine the list in which the user made a purchase. The final methodology to build the data collection and labeling is depicted in Figure~\ref{fig:training_process}.

\subsection{Evaluation}

For offline evaluation, we use accuracy on our unbiased sampled dataset as the primary metric because the task essentially involves selecting the correct collection from two options. Each collection is scored independently, and the prediction hinges on which collection receives the higher score.

Preliminary observations of our A/B test results have shown that offline gains in accuracy translate reliably into online conversion lifts, allowing us to forecast the expected impact of any model change before deployment. This finding is very different from what was reported in other applications \cite{bernardi2019150}, reflecting the quality and importance of our unbiased dataset.

\section{RESULTS}

Along the course of 12 months, we conducted online A/B experiments with a fraction varying between 5 to 10\% of iFood's user base with the duration of 2 to 4 weeks each, testing incremental improvements built upon each other. The measured metrics were the Card Conversion Rate (CCR), Home Conversion Rate (HCR) and App Conversion Rate (APC), where conversion is defined as the user placing an order. The results are shown in Table~\ref{tab:red_results}.

Our experiments were conducted in the following temporal sequence: baseline $\to$ RED v1 $\to$ RED v2 $\to$ RED v3 $\to$ RED v4 $\to$ RED v5. The baseline consists of recommending the most popular collections. RED v1 is our first model version, trained with the dataset described in Section~\ref{first_data}. RED v2 added monotonicity constraints as described in Section~\ref{sec:controlability} and migrated embeddings calculated with FastText to a two-tower model optimized for search retrieval, described in Sections~\ref{sec:user_embedding} and ~\ref{sec:collection_embedding}. RED v3 added information about collection popularity measured as their average embedding similarity with the users in the meal shift. RED v4 migrated the training and evaluation dataset as described in Section \ref{sec:mature_data}. RED v5 introduced a hand-engineered similarity features, defined as the number of user orders in the collection restaurants in the meal shift normalized by the number of restaurants in the collection.

Taken together, the experiments achieved a \textbf{97\% uplift} in card conversion rate, 
a \textbf{4.5\% increase} in home‐page conversion, and a \textbf{1.4\% rise} in overall app conversion. 
This improvement in card performance is also reflected over time, where we observed a 96\% increase in orders originating from RED cards between March 2024 and March 2025. 
In a final A/B test—after all enhancements had been deployed—
we confirmed a statistically significant \textbf{0.4\% uplift} in total app orders.

An interesting finding from our experiments was that our offline accuracy metrics closely correlated with the online A/B test results. This preliminary observation suggests that the unbiased nature of our dataset makes it a good proxy for online conversion. This relationship is illustrated in Table~\ref{tab:red_offline_results}, which presents the online and offline improvements from our three most recent experiments.

Given the successful online results, RED recommendations were deployed to 100\% of the user base, becoming an integral part of the main app variant.

\begin{table}[H]
  \centering
  % \scriptsize
  % \setlength\tabcolsep{3pt}
  \caption{Relationship between offline accuracy improvement in percentage points and online A/B test results in the three most recent experiments. Offline performance closely aligned with the observed online A/B test lift.}
  \label{tab:red_offline_results}
  % \resizebox{\linewidth}{!}{%
    \begin{tabular}{cccc}
      \toprule
      Control & Variant & Offline Accuracy Diff & CCR  \\
      \midrule
      RED v2 & RED v3
        & +3.20\% & +11.51\% \\
      RED v3 & RED v4
        & +5.50\% & +18.32\% \\
      RED v4 & RED v5
        & +2.22\% & +6.30\% \\
      \bottomrule
    \end{tabular}%
  % }
\end{table}

\section{FUTURE WORK}

The plans for future work can be grouped into three areas of improvement: (1) incorporate user-session signals, (2) develop a sequential model, and (3) enhance the training dataset.

There are two hypotheses underpinning the use of user-session signals to refine recommendations. The first is that these signals are especially valuable for new users, who have a limited history on the platform and thus receive only weakly personalized suggestions. We believe that real-time browsing data captures these users’ immediate preferences, enabling us to deliver personalized recommendations even on their very first order. The second hypothesis is that there is an upper bound on the accuracy an offline (D-1) model can achieve when predicting a user’s next meal. To illustrate this ceiling: even people themselves struggle to foresee precisely what they will want to eat next weekend, wavering among two or three dish options. However, a few minutes before mealtime, the searches and clicks a user makes on the platform reveal their exact intention, and we can leverage those signals to deliver far more accurate recommendations. In preliminary experiments that used the user’s most recent search in real time, we achieved highly positive results.

We also see opportunities in employing sequential models to generate recommendations. We believe sequence information is crucial in meal recommendation—many people prefer to rotate cuisines over time. Moreover, sequential representations have proven to be a highly effective way to learn user and item embeddings, which can be reused by other applications within the company.

Regarding the training data, we’ve identified two avenues for evolution. One is to leverage the full set of impression-and-purchase data in tandem with our sampled records, for example, by training in stages and finishing with a fine-tuning phase on an unbiased dataset. Another is to adopt a more efficient sampling strategy than uniform sampling—sampling more frequently from collections that convert better at a given time of day.

\section{CONCLUSION}

In this work, we presented RED, a scalable and automated recommendation system for personalized collections of food items and restaurants on iFood, Latin America’s leading food delivery platform. To the best of our knowledge, this is \textbf{the first work} detailing the recommendation of curated collections of dishes and restaurants in the food delivery domain, addressing the unique challenges of a highly dynamic and complex environment. Our approach leverages content-based representations, user–collection similarity metrics, and contextual features within a LightGBM framework, enabling effective generalization to previously unseen collections. We addressed critical challenges such as data scarcity and visibility bias by bootstrapping from existing category interactions and systematically collecting unbiased production data.

Our iterative experimentation demonstrated significant improvements: a 97\% uplift in Card Conversion Rate, 4.5\% increase in Homepage Conversions, and a 1.4\% rise in overall App Conversion. Notably, we made a promising preliminary observation that improvements in offline accuracy closely tracked with gains in online conversion metrics, suggesting the robustness of our evaluation pipeline and dataset.

This research represents a foundational step in scaling personalized collection recommendations in dynamic two-sided marketplaces. Our methodological contributions—such as embedding-based representations, monotonicity constraints to prevent overfitting, and a rigorously curated dataset—can serve as a blueprint for similar recommendation challenges in other domains.

Looking forward, we anticipate that incorporating real-time session signals and sequential modeling will further enhance personalization quality. Additionally, refining our dataset with more sophisticated sampling techniques offers a promising avenue for sustaining model improvements. Ultimately, RED’s deployment to iFood’s entire user base illustrates the tangible business impact of machine learning solutions grounded in both rigorous experimentation and practical constraints.

% \begin{acks}
% This work is supported by iFood.
% \end{acks}

\bibliographystyle{ACM-Reference-Format}
\bibliography{main}

\appendix

\section{SUPPLEMENTARY INFORMATION}

\textbf{Content Management System (CMS)}: iFood’s content management interface. 
It is a platform used to create and manage homepages, e.g. create collections, add collections to homepages and define to which segment of users the collections can be rendered.

\end{document}